# Cornu's spiral in the Fresnel regime studied using ultrasound: A phase study

Akira Hitachi
Molecular Biophysics, Kochi Medical School, Nankoku, Kochi 783-8505, Japan
Electronic mail: jm-hitachia@kochi-u.ac.jp

**Abstract.** A simple experimental technique for measuring the phase and amplitude of diffracting ultrasound wave [A. Hitachi and M. Takata, Am. J. Phys. **78**, 678 (2010)] has been applied to diffracting objects with straight edges as a demonstration of the Cornu spiral. Babinet's principle is studied observing the ultrasound field behind a slit and a complementary strip obstacle and is verified directly by comparing vectors (phasors) in the complex plane. The phase of the diffracted wave observed in the geometrical shadow of the straight screen has the form of a cylindrical wave originating at the edge of the straight screen as the boundary diffraction wave proposed by Young. In addition, the incident wave has a phase delay of $\pi/4$ behind the wave passing through on the center line of the slit, the plane of symmetry, has been observed as predicted by Huygens-Fresnel diffraction theory.

## I. INTRODUCTION

The diffraction by an opaque screen with a straight edge is the simplest application of diffraction theories and has been studied extensively.[1,2] The Cornu spiral, a plot of the Fresnel integrals, has been used to introduce the concept of diffraction in optics courses and to explain the diffraction patterns in the Fresnel regime.[1-6] The Cornu spiral can give a clear picture of diffraction and the relation between the fields such as the field behind a straight slit $U_1$ and a strip obstacle $U_2$. When a slit and a strip are complementary to each other, i.e., the opening in one corresponds exactly to the opaque portion of the other and vice versa, Babinet's principle (1837) is expressed as,[1-7]

$$U_0 = U_1 + U_2, \qquad (1)$$

where $U_0$ is the field observed without the diffracting objects. The Cornu spiral and Babinet's principle are some of the most difficult diffraction topics in the student laboratory because the intensity measurements require complex calculations to compare with theory. A simultaneous measurement of the amplitude and phase enables one to study Babinet's principle directly, as has been demonstrated recently by an experiment by Hitachi and Takata using ultrasound.[8] In that experiment the diffracted wave are detected with a piezo-ceramic transducer and the signal is observed by an oscilloscope. The essence of the diffraction theory can be seen by simply drawing vectors (phasors), using a graduated ruler and a protractor, in the complex plane with the Cornu spiral printed.

The rigorous solution of Sommerfeld diffraction theory[1,2] shows that the field observed in the geometrical shadow of the straight screen has a form of a cylindrical wave originating at the edge of the screen. Young's approach explains diffraction as an interference between the incident wave $U_0$ and the boundary diffraction wave $U^B$ originating at the edge of the diffracting object. The observation of the cylindrical wave strengthens the idea of a boundary diffraction wave in Young's approach.[9] The intensity in the geometrical shadow of a half plane decreases monotonically as the observation point $P$ goes into the shadow. It is not possible to obtain the phase by the intensity measurements, whereas the direct measurement of the phase is possible by the present method. In the course of this study, a difference of about $\pi/4$ between the phase and path in Young's approach has been observed behind the central axis of a straight strip obstacle and is discussed later in the paper.

The Cornu spiral is tilted $\pi/4$ to the real axis in the complex plane.[3] The phase of the incident wave $U_0$ is $\pi/4$, or one-eighth period behind that of the wave passing through on the central line, the centre of the linear-zone system. This curious feature also occurs in the treatment of the diffraction in the circular zones, a delay of $\pi/2$, and was regarded as a defect of Huygens-Fresnel diffraction theory,[3,5] or was accounted for by assuming that Huygens' secondary wavelets oscillate a quarter of period out of phase.[2] However, the phase delay of $\pi/2$ has been confirmed in the diffraction of ultrasound by a small circular aperture.[8] This subject is discussed in Refs. 3 and 5, and more in Ref. 8, therefore in this paper, the experimental evidence is presented but only briefly discussed.

As a demonstration of Fresnel diffraction, the diffraction of ultrasound by a slit and a complementary opaque strip is observed, and the Cornu spiral as well as Babinet's principle is studied further as presented below. The amplitude and phase behind a half plane and a narrow slit with the width close to the wavelength $\lambda$ being used are also measured and discussed in relation to the Cornu spiral. The results are also discussed with respect to the edge diffraction approach by Young. The terms and mathematical expressions of optics are used in this article because most of the original works have been done for light diffraction.

**II. THEORETICAL BACKGROUND**

Two different approaches for the diffraction: Huygens-Fresnel diffraction theory (1818) and Young's theory (1802), are treated in the following sections. In Huygens-Fresnel diffraction theory the diffraction is described as the sum of all the contributions from every point in the diffracting aperture and the edge of the diffracting object plays no role, except defining the area contributing. Young assumed that the incident wave undergoes a kind of reflection at the edge of the diffracting object and interpreted the diffraction as the interference between this boundary diffraction wave and the incident wave. Huygens-Fresnel theory was successful and was given a mathematical foundation by Kirchhoff (1882), and refined by Sommerfeld (1894). Young's approach was forgotten for about a century. However, Young's approach is intrinsically simple and physically appealing. The superscript B appears on $U$ denotes the boundary diffraction wave in the following sections.

**A. Graphical treatment of diffraction: The Cornu spiral**

Consider diffraction of a plane wave by a semi-infinite plane obstacle (a half plane) on the *x-y* plane bounded by a sharp straight edge as shown in Fig. 1. The wave is incident normal to the half plane along the *z* axis. The observation point $P$ is at distance $z$ from the half plane and on the edge of the geometrical shadow. The problem becomes independent of *y* and can be reduced to two-dimensions.

The half-period (linear) zones can be defined by dividing a plane wave front at the diffraction plane such that the distances from the edges ($M_i$) are successively one-half wavelength farther from $P$. The areas of the half-period zones decrease rapidly as $s$, the distance from the edge of the screen $M_0$ on the diffraction plane, increases. The calculation with the linear zone is not as useful as that with a circular zone since the evaluation of the contribution of each zone is not simple.

The diffraction of a half plane can be explained by introducing the Cornu spiral. Clear descriptions of the Cornu spiral can be found in a number of optics books.[1-6] The contribution of wavelets from the diffraction plane may be expressed using Fresnel integrals, $C(v)$ and $S(v)$, where a dimensionless variable $v$ is given by,[1-5]

$$v = s\sqrt{\frac{2}{z\lambda}}. \qquad (2)$$

The field observed at $P$ due to the contribution from a region of 0 to $v$ is given, for an amplitude $A_0$ of the incident wave, as follows:[1,4]

$$A_0 \exp(-i\omega t)[C(v) + iS(v)] = A_0 \exp(-i\omega t) \int_0^v \exp(\frac{1}{2}i\pi u^2)du. \qquad (3)$$

The numerical values of the integrals are given elsewhere[3,5] or implemented in software for mathematics. The Fresnel integrals are often expressed in a separated form of real and imaginary (cosine and sine) parts. They can be reduced to a form in Eq. (3) by Euler's identity.[1,4] The behavior of the Fresnel's integrals is illustrated by means of a graphical representation of the Cornu spiral as shown in Fig. 2. $A_0$ is normalized to be $\sqrt{2}$. The variable $v$ is the distance measured from the origin along the spiral. The simple harmonic wave is presented as the projection on the coordinate of the real part. The figure applies for $t=0$. The phasors rotate clockwise with an angular velocity $\omega$ following the convention in optics of representing an advance of time. The $v$ values of the edge of the half-period zones are given by $(2n)^{1/2}$ where $n$ is 1, 2, … and are marked by diamonds on the spiral. The amplitude and phase of the diffracted wave for a slit shown in Fig. 1 are given by a phasor $U_1$, a chord of the arc from $O$ to $B$. Fresnel integrals of a half plane are obtained by integration of Eq. (3) from 0 to $\infty$ and give a value of $(1+i)/2$ and represented by a phasor $OQ$.

The phasor for a slit set along the $s$ axis is given by Eq. (3) setting $t = 0$, specified by the limits $v'$ and $v$ for two edges,[4]

$$A_0 \int_{v'}^{v} \exp(\frac{1}{2}i\pi u^2)du = A\exp(i\phi), \qquad (4)$$

and is presented by the straight line $I'J$ connecting the two points on the Cornu spiral as shown in Fig. 3. The amplitude $A$ is the length of the line. The phase $\phi$ is the angle the line makes with the real axis.

It is practical to prepare the Cornu spiral with $v$ values indicated to show the theoretical amplitude and phase for a slit placed on the $s$ axis. The horizontal lines in Fig. 4 show the positions of slits and a strip placed on the $s$ axis. The values for the start $v'$ and stop $v$ points on the Cornu spiral are calculated by Eq. (2) with $s$ values for the position of edges of the slit. Then the phasor for the slit in Fig. 4 is obtained by connecting the two points as shown in Fig. 3.

A curious feature is that the phase appearing in Eq. (4) is $\phi$, not $\theta$ which is the phase relative to the phase of the incident wave. The real axis is the phase of the wave passing though the center of the zone system ($OP$ in Fig.1) and that should be the same as the incident wave. It is usually only the diffracted intensity, $|Ae^{i\phi}|^2 = A^2$, that is of interest in optics, in which case only the length $A$ is needed; therefore it has not drawn much attention. The difference of the phases is not obtained from the intensity measurements. The phase delay may be partly explained in that Huygens-Fresnel diffraction theory describes the field observed at $P$ as the results of contributions of the waves passing through all the space; consequently, the phase for $U_0$ can be different from the wave passing through the center of the zone system. It is not reasonable to assume that Huygens' wavelets oscillate $\pi/2$ out of phase in diffraction of the circular geometry and $\pi/4$ of the straight geometry.

When $P$ moves on the $x$ axis, the zone system follows. The same result is obtained by moving $P$ or moving the diffracting object in the opposite direction. The contribution of any given portion of the wave front may be obtained from Eq. (4) with the appropriate limits of the integral, or graphically, the chord of the segment of arc defined by a proper choice of $v'$ and $v$.

The above argument can be applied for a cylindrical incident wave. The value of $v$ is then given as

$$v = s\sqrt{\frac{2(z_0 + z)}{z_0 z \lambda}}, \qquad (5)$$

where $z_0$ is the distance from the source to the diffracting plane. However, parallax has to be taken into account when $P$ is off the axis ($x \neq 0$).

### B. Young's approach: The boundary diffraction wave

An edge diffraction approach by Young describes the field $U_L$ in the geometrical shadow of a half plane $L$ as the boundary diffraction wave $U_L^B$ (the cylindrical wave originating at the edge of the screen) and, in the direct beam, as the superposition of this wave and the incident wave $U_0$.[1,2,4,5]

$$U_L(x>0) = U_0 + U_L^B(x>0) \quad (P \text{ in the direct beam}), \tag{6}$$

$$U_L(x<0) = \phantom{U_0 +} U_L^B(x<0) \quad (P \text{ in the geometrical shadow}). \tag{7}$$

With the complementary plane $R$, Eq. (6) and Babinet's principle, $U_0 = U_L + U_R$, yield $U_R(x>0) = -U_L^B(x>0)$. This equation together with one corresponding to Eq. (7), $U_R(x>0) = U_R^B(x>0)$, leads to a relation $U_L^B = -U_R^B$, then $U^B(-x) = -U^B(x)$ by considering symmetry. This relation requires a change of $\pi$ in the phase of the boundary diffraction wave across the edge of the geometrical shadow. Fig. 5 illustrates the idea of Young's approach. The incident wave $U_0$ disappears suddenly at the straight edge of the geometrical shadow and the discontinuity is compensated by the boundary wave $U^B$. The total diffracted field must be continuous across the edge of the geometrical shadow; $U_L^B(-0) = U_L^B(+0) + U_0$. Consequently, the amplitude $A_{Hf}$ at the edge of the geometrical shadow is one half that of the incident wave and the phase $\theta_{Hf}$ is the same as the incident wave.

The existence of the Poisson's bright spot behind a circular disk can be explained as follows: $U^B$ interact constructively since the distances between every point on the rim of a circular disk to the axis are the same. On the other hand, it is not simple to explain why the intensity of the spot $U^B$ is the same as that of $U_0$. The phase of $U^B$ cannot be obtained from the intensity measurements since the bright spot is always there with the same intensity. However, the measurement including the phase shows $U^B \approx -U_0$ when the disk contains the first circular zone, in accordance with the path difference.[8] In contrast, the amplitude (or the intensity) observed for a circular aperture on the axis shows a maximum when the aperture contains the first circular zone; $U_0$ and $U^B$ interact constructively, while the path difference for the two waves is $\lambda/2$. $U^B$ should be $\pi$ phase shifted. A zone plate, a circular aperture made by blacked out even-numbered (alternatively, odd-numbered) circular zones acts as a lens and can obtain a high intensity at the focal point on the central axis. The zone plate may be explained if $U^B$ is $\pi$ phase shifted at the rim when diffracted into the beam and is not shifted when goes into the geometrical shadow. The path increases $\lambda/2$ as successive zones is included, yet every contribution from the rim interacts constructively.

The division of the diffracted wave into the incident wave and another wave does not necessarily provide evidence of the boundary diffraction wave. In fact the field behind a diffracting object can be divided into two terms; the incident wave and the wave due to a complementary object. However, the incident wave in the geometrical shadow is not a physical reality. Maggi (1888) and Rubinowicz (1917) showed that the Kirchhoff diffraction integral can be divided into the incident wave and the boundary wave, which is a line integral over the rim of an aperture, according to Young's approach.[1,2,4] The condition on the boundary diffraction wave is analogous to the laws of reflection, $U^B$ changes by a phase of $\pi$ across the edge of geometrical shadow. Sommerfeld obtained a rigorous solution for the diffraction of the plane wave by a half plane (1894).[1,2] The solution shows that the light propagates in the form of a cylindrical wave which appears to proceed from the edge of the half plane in the geometrical shadow. In the beam it is represented as superposition of the cylindrical wave and the incident wave. The physical reality of the boundary diffraction wave has been demonstrated by Ganci at the straight edge using Young's double slit.[10] The edge diffraction approach has not been proved beyond the limits of validity of Kirchhoff theory, when the wavelength is considerably small compared to the diffraction opening.[1,2,4,9] Young's approach is also referred as the geometrical theory of diffraction[11] and has been applied in acoustics such as seismic field experiments.[12]

It has been pointed out that the position of extrema in the fringe intensity pattern behind a half plane requires an additional optical path of about $\lambda/8$ for the boundary wave $U^B$ compared to the incident wave.[13,14] The extra phase of about $-\pi/4$ was attributed to an effect similar to the Gouy effect, in which the phase changes by $\pi/4$ in crossing the focus. The disagreement for the optical paths for $U_0$ and $U^B$ can be seen in the Cornu spiral. The direction of $U_0$ does not align with a line that goes through the edges of the zones; they are about $\pi/4$ off (see Fig. 2).[15] The off-alignment, however, does not occur in the vibration spiral in the circular symmetry; $U_0$ and the edges of the zones lie on the imaginary axis. The phases of the wave behind a circular aperture and a disk observed on the central axis are in accordance with the phase of $U^B$ as the same as $U_0$ and a $\pi$ phase difference for the circular disk and aperture, respectively, on the rim.[8] No extra phase of $-\pi/4$ nor $-\pi/2$ is necessary.

**III. EXPERIMENTAL METHOD**

The apparatus used is shown schematically in Fig. 6; the ultrasound instrument is basically the same as that described before.[8] The piezo-ceramic transducers were used as the transmitter and the receiver of ultrasound. The transmitter was driven by an oscillator (Agilent 3321A) and the signal from the receiver was observed by an oscilloscope. A 25.4 kHz frequency for the transmitter (Nicera T2516A1) and receiver (Nicera R2516A1) is used in addition to a 40 kHz frequency in the present measurement. The wavelengths at room temperature were 13.5 mm and 8.6 mm for 25.4 kHz and 40 kHz, respectively. The signal transmitting and accepting angles of the 25.4 kHz frequency were 85° FWHM and the diameters of the transducers were 7 mm. Those of the 40 kHz frequency were 50° FWHM and 95° FWHM, and 7 mm and 6 mm for the transmitter (T4016) and receiver (R4010A1), respectively. The 25.4 kHz frequency was used for small $v$ and the 40 kHz frequency was used for a wide range of $v$.

The signal from the receiver was put into channel 1 of the oscilloscope and the voltage $V_{Ref}$ applied to the transmitter was monitored in channel 2 and used as a reference signal. The phase $\theta'$ was obtained by $\theta' = 2\pi\tau/T$ where $\tau$ is the delay of the signal with respect to $V_{Ref}$ and $T$ is the period of the ultrasound as shown in Fig. 7. Then the $\theta$ values were obtained with respect to the phase of the incident wave $U_0$ at $x=0$. The voltage used was typically 6 $V_{p-p}$. Since the signals can be quite small, co-axial cables were used. Some transducers have a small output at 90° or more. That output can make serious distortions of the sound pressure in front. The transducer should be separated from its support whenever such distortion is observed. Sheets of synthetic sponge were used to avoid the reflection from the desk when necessary.

It is not easy to arrange a coherent plane wave or cylindrical wave, which is suitable for the investigation of the diffraction that involves a straight edge. A coherent quasi-cylindrical wave was prepared by simply setting the distance $z_0+z$ and the height $2h$ of the edges much longer than $z$, $s$ and $x$. The distance, $z_0+z$, was set to be 1080 mm (80$\lambda$ at 25.4 kHz and 126$\lambda$ at 40 kHz). The height $h$ of the transmitter and the receiver was 340 mm.

A picture of the apparatus is shown in Fig. 8. Two screens and a strip were made. A straight blade was made from 0.4 mm (360 g/m$^2$) thick Kent paper (high quality, high density, smooth surface, and stiff paper) of 210 mm width and 680 mm long which was held along length and glued to enhance strength; a 3 mm width was left as a single thickness. The blades were set on a pair of screens of 450 mm width and 680 mm high cork boards. The paper blade extended from the edge of the cork board by 15 mm (> 1$\lambda$). In addition, a straight strip obstacle of 72 mm width and 680 mm high with both blades was prepared in similar way and was put on a cork strip board of 44 mm width and 680 mm high. Accurate alignment is essential in the present method. The blades can be adjusted and replaced and the apparatus was put on a 5 cm-square-ruled mat.

**A. Babinet's principle at straight edges**

An example of the configuration for the study of Babinet's principle using two screens and a strip is shown schematically in Fig. 4. The receiver was set at $x=0$. Solid horizontal lines show screens and/or strip placed on the $s$-axis. Positions of edges are shown in $s$ ($\lambda$) and $v$ is calculated by Eq. (2). The value of $z$ is assumed to be 10$\lambda$ and the positions of edges $s$ are taken to be ±4$\lambda$ in Fig. 4. $Q'$ and $Q$ represent $v = -\infty$ and $\infty$, respectively. In Fig 4 (a) represents without the diffracting object (the incident wave) and (f) shows a half plane. The phasors given by Eqs. (3) and (4) are shown as arrows in Fig. 3.

Two types of Babinet's principle can be considered with the straight edges. One involves two complementary screens ((b) and (c)), and the other involves a slit (d) and a strip (e). The fields $U_{SL}$ and $U_{SR}$ behind complementary screens $S_L$ (b) and $S_R$ (c) are given by Eq. (4), with limits of integration $v$ to $\infty$ and $-\infty$ to $v$, respectively. Phasors $U_0$, $U_{SL}$ and $U_{SR}$ are expressed as $Q'Q$, $BQ$ and $Q'B$, respectively, as shown in Fig. 3. Babinet's principle, $U_0 = U_{SL} + U_{SR}$, was studied.

The set of screens and a strip were placed at the centre of the slit or the strip on the plane of symmetry. The field $U_{ap}$ behind a slit (d) is given by Eq. (4), with limits of integration $-v$ and $v$. The limits of integration for the field $U_{ob}$ behind a strip (e) are $-\infty$ to $-v$ and $v$ to $\infty$. Babinet's principle is $U_0 = U_{ap} + U_{ob}$; phasors $U_0$, $U_{ap}$ and $U_{ob}$ are expressed as $Q'Q$, $B'B$ and $Q'B'+BQ = OC$, respectively, as shown in Fig. 3. The relation of the phasors can be shown in the first quadrant as $OQ$, $OB$ and $BQ$, because of the symmetry of the positions.

Use of the symmetry plane allows one to study Babinet's principle, $U_0 = U_{ap} + U_{ob}$, without a strip as follows. One screen (screen 1) is place with its edge on the plane of symmetry and acts as a half plane (f) (Figs. 1 and 4). The wave $U_{0/2}$ is measured. Then the other screen (screen 2) is placed with its edge separated by a distance $s$ from the half plane to produce a straight slit (g). $U_{ap/2}$ was measured. Next, screen 2 was removed and screen 1 was moved so that the edge was set at the place where the edge of screen 2 was placed (h). $U_{ob/2}$ was measured. One obtains the half field version, $U_{0/2} = U_{ap/2} + U_{ob/2}$.

The above description gives the ways of setting the diffracting objects, and the positions $B'$, $B$, $z_0$ and $z$ are different for each of the measurements described in Results and Discussion.

**B. The amplitude and the phase patterns behind diffracting objects**

The receiver was set on a rail placed perpendicularly to $z$ axis and the distribution of the signal was scanned along the $x$ axis beyond the half plane (Fig. 6). The amplitude $A$ and the phase $\theta$ were registered as a function of $x$. The patterns due to a slit and a strip of 72 mm wide were also measured. The measurements were performed at 40 kHz and the distance $z$ between the diffracting object to the scanning line was 120 mm. The relevant Cornu spiral is in the first and the third quadrants.

**C. The π/4 phase shift**

Narrow slits, with a width close to the wavelength λ were set up, these contain only a small fraction of a half-period zone and were used to investigate the phase delay $\Delta\theta$ of the incident wave with respect to the wave from the centre of the zone system. The field behind the narrow slit $U_{na}$ was measured at $x=0$ and compared with the Cornu spiral. The amplitude $A_{na}$ and the phase $\theta_{na}$ patterns were also observed.

**IV. RESULTS AND DISCUSSION**

The relation of phasors observed at $x=0$ for various configurations of screens and a strip are shown in Fig. 9. The frequency used was 25.4 kHz except for a set of data for $U_{ap}$ and $U_{ob}$, which was taken at 40 kHz. The phase $\theta_0$ for no diffracting object was taken to be $\pi/4$, following the convention of representing the Cornu spiral. The amplitude $A_0$ was normalized to be $10\sqrt{2}$. Babinet's principle is rewritten as $U_{ap} = U_0 - U_{ob}$, and the endpoints for two sets of a half portion, $U_{ap}/2 = U_0/2 - U_{ob}/2$, were plotted (○ and ●, respectively for $U_{ap}$ and $U_{ob}$). In this plot, measured $U_{ap}$ and $U_{ob}$ can be compared with the Cornu spiral separately. The width $2s$ was 72 mm, and $z=135$ and 120 mm, respectively for 25.4 kHz and 40 kHz. In addition, a set of $U_{SR} = U_0 - U_{SL}$ (△ and ■) and a set of $U_{ap/2} = U_{0/2} - U_{ob/2}$ (□ and ■) for $s = 22$ mm at $z=135$ mm are shown. The starting point for $U_{SR}$ (△) is $Q'$ in the third quadrant as shown in Fig. 3. The letters in brackets show the configuration as in Fig. 4. Five measurements are performed for each setting; the plotted points show the results. The main contribution to the measurement error comes from errors in the phase and in alignment. The bars on the Cornu spiral show $v$ values calculated for each setup for the cylindrical incident wave (Eq. (5)). The results compare reasonably well with the Cornu spiral and verified Babinet's principle.

The results obtained for narrow slits of 14 mm (~1λ) and 27 mm (≈2λ) width placed at $z=235$ mm taken at 25.4 kHz are also plotted (◊) on the complex plane (Fig. 9) and they lay well on the Cornu spiral. The anomalous orientation of the Cornu spiral, at an angle of $\pi/4$ to the real axes was detected.

The result obtained for the half screen placed at $z=120$ mm (14λ at 40 kHz) as a function of the distance $x$ from the geometrical edge is shown in Fig. 10. The amplitude $A_{Hf}$ pattern shows a structure described by the Cornu spiral.[1-6] On the edge of the geometrical shadow $A_{Hf} = A_0/2$. The same result for moving $P$ on the x-axis is basically obtained by moving the half plane in the opposite direction in the $s$ axis as described in Sec. II.A. The phasor for the half plane is given by Eq. (4) specified by the limits $v$ and $\infty$. As the $x$ value increases in the beam ($x>0$) the $s$ value decreases in $s<0$, the contribution from $v<0$ increases. The end point of phasor $U_{Hf}$ is on $Q$, and the starting point moves from $O$ into the third quadrant $H'$, then rotates around $Q'$ with diminishing radius as $x$ increases. In the beam, $A_{Hf}$ oscillates with diminishing amplitudes as $x$ increases and approaches asymptotically to the value of $A_0$. In the shadow ($x<0$), the half plane covers small values of $v$, the end point of phasor is $O$ and the starting point moves on the Cornu spiral on the first quadrant towards $O$. The value of $A_{Hf}$ decreases towards zero as $P$ goes into the geometrical shadow.

    The observed phases are also shown in Fig. 10. The observed phase $\theta_0$ for the incident wave lay well on the solid parabolic curve which shows the phase $2\pi[R_x-(z_0+z)]/\lambda$, where $R_x$ is the distance between the transmitter and the receiver, of the spherical wave originating at the transmitter. The observed phase $\theta_{Hf}$ for a half screen shows features also described by the Cornu spiral (Fig. 3). The angle $\Delta\theta_{Hf}$ that phasors $U_{Hf}$ ($H'Q$) and $U_0$ ($Q'Q$) make gives deviation of $\theta_{Hf}$ from the incident beam $\theta_0$. Since the length of the phasor is much larger than the radius, the phase $\theta_{Hf}$ stays within about +8° and -16° from $\theta_0$. The angles can be easily obtained by measuring, with a protractor, the angles which the line $Q'Q$ and the tangential lines to the Cornu spiral on the third quadrant drawn from $Q$ makes. The deviation oscillates with diminishing amplitude as $x$ increases and $\theta_{Hf}$ approaches asymptotically to $\theta_0$. On the other hand, the phase change is large in the shadow; the Cornu spiral indicates a monotonic rotation of the $U_{Hf}$ phasor shown by $BQ$ (in Fig. 3). The broken curve in Fig. 10 shows the calculated phase delay $2\pi(r-z)/\lambda$ (Fig. 6), where $r$ is the distance between the edge of the half plane to the receiver, for the cylindrical wave originating at the edge of the screen. The curve can be regarded as an envelope of Huygens' spherical wavelets or the boundary diffraction wave in Young's approach.

    The amplitude and phase obtained for a slit and a strip of 72 mm width at 40 kHz are shown in Fig. 11. The vertical broken lines show the position of the edges of geometrical shadows. The amplitude patterns appear as expected from the Cornu spiral.[3,5] The amplitude for the slit $A_{ap}$ at $x=0$ in the present setup shows a minimum surrounded by two maxima. A bright line, which corresponds to Poisson's bright spot behind a circular disk, appears behind a strip along the central line parallel to the strip. The amplitude $A_{ob}$ at $x=0$ of the bright line is smaller than $A_0$ and decreases as the width of the strip increases. The amplitude pattern $A_{ob}$ behind a strip can be explained as follows. When the observation point $P$ goes off the axis, the phasors $U_{obR}$ and $U_{obL}$, the contributions from each side of the strip, shift to opposite directions as shown by $EQ$ and $Q'F'$ in Fig. 3. A minimum is observed when two phases becomes opposite to each other. The distance from the edge increases further, $U_{obR}$ and $U_{obL}$ rotate then they come to a point where the phases coincide and give a maximum. This process continues till $P$ comes close to the geometrical edge. Therefore the centre is a maximum surrounded by

minima and maxima.

The phases for the slit $\theta_{ap}$ and strip $\theta_{ob}$ observed outside the edges show similar features as observed in the half plane. The phase $\theta_{ob}$ in the beam basically follows the parabolic curves representing the phase of the incident wave. The phase $\theta_{ap}$ inside the geometrical shadow follow the broken parabolic curves representing the cylindrical wave originating at the edges of the slit. The phase for the cylindrical wave is calculated as $2\pi(r-z)/\lambda+\Delta\theta_{ed}$, where $\Delta\theta_{ed}=28°$ is the phase delay for the incident wave arriving at the edge of the slit to that arriving at the centre of the slit. The phase $\theta_{ap}$ in the beam is ahead of $\theta_0$ except in the vicinity of the centre. The phase $\theta_{ob}$ inside the shadow is behind $\theta_0$, and $\theta_{ob}$ moves backward, step-like, as $P$ goes into the shadow. The phase $\theta_{ob}$ stays constant around $A_{ob}$ maxima and changes quickly at the $A_{ob}$ minima. The width of the slit can be obtained from the value of $\theta_{ob}$ at $x=0$.

The phase $\theta_{na}$ pattern obtained for a 14 mm (~1$\lambda$ at 25.4 kHz) and a 12 mm (1.4$\lambda$ at 40 kHz) slits placed at $z=$ 235 and 220 mm, respectively, are shown in Fig. 12. The vertical lines show the edges of the geometrical shadow. The broken and chain parabolic curves show calculated phase differences $2\pi(r-z)/\lambda$, where $r$ is the distance between the centre of the narrow slit to the receiver, with an added phase difference of -45°. The observed phases lay on the curves. The results verify that the wave passing through the centre of the linear-zone system is $\pi/4$ ahead of the incident wave, as predicted by the representation of Fresnel diffraction by the Cornu spiral.

The incident wave is on a line tilted $\pi/4$ in the Cornu spiral and it is on the imaginary axis ($\pi/2$) in the treatment of circular zones. However the difference of $\pi/4$ is consistent within Huygens-Fresnel diffraction theory. The wave coming from the centre of the linear-zone system in the Cornu spiral can be regarded as a wave from one dimensional slit. The same argument in Sec. II.A can be applied by putting the slit on the $s$ axis in Fig. 1, therefore it gives an additional $\pi/4$ delay for the Cornu spiral. As a result, the incident wave has a phase delay $\pi/2$ and lies on the imaginary axis.

The phase relations observed in the present measurements are reviewed below from Young's approach. The field $U_{Hf}$ observed in the geometrical shadow of a half plane shows the cylindrical wave originating at the edge of the screen and can be regarded as the boundary diffraction wave $U^B$. At the boundary into the shadow, $U^B$ is observed to be in phase with $U_0$. On the other hand in the beam, $U^B$ cannot be separated from $U_0$. The amplitude $A_{Hf}$ observed at the edge of the geometrical shadow is a half that of the incident wave as shown in Fig. 10, therefore, $U^B$ changes in phase by $\pi$ across the boundary as discussed in Sec. II. B. The rigorous solution using Sommerfeld diffraction theory also shows the diffracted waves have a form of $U^B(-x)= -U^B(x)$, a $\pi$ phase difference.[7] The extra phase for $\theta_{Hf}$ of $\sim\pi/4$ on the edge of the geometrical shadow as discussed in Sec. II.B was not detected at $x=0$.

The field $U_{ap}$ in the geometrical shadow of the slit can be expressed by two boundary diffraction waves; one comes from the near edge; the phase at the edge is the same as the incident wave. The other comes from the far side; the phase is reversed at the edge. The amplitude for the wave from the near edge is much larger than the other, therefore coarse parts of the amplitude $A_{ap}$ and phase $\theta_{ap}$ are due to $U^B$ from the near edge, and fine parts are due to $U^B$ from the far edge.

A curious feature relating to that discussed in the last paragraph in Sec. II.B appears in the phase of $U_{ob}$ ($2KQ$ in Fig. 9) observed at $x=0$ at 25.4 kHz. The phasor $-U_{ob}$ aligns with $U_0$ and the phase $\theta_{ob}$ differs by $\pi$ from $\theta_0$ as shown in Fig. 9. The difference of the path from the source $S$ to $P$ between via the center and via the edge of the strip is not close to $\lambda/2$ as the measured points ● are about 40° away from $M_1$. In Young's approach, $U_{ob} = 2U_{ob}^B$ on the central axis (Eq. (7); a factor 2 accounts for the boundary diffraction wave $U_{ob}^B$ coming from both sides), therefore the path should be $\lambda/2$ or an extra phase of about $-\pi/4$ mentioned earlier might be needed. However, the phase $\theta_{Hf}$ of the wave into the shadow does not show the extra phase (Fig. 10). The disagreement between the phase and the path is intrinsic in the Cornu spiral treatment. It would occur also for the incident plane wave, and within the approximation of the Kirchhoff theory, where Young's approach has been proven to be valid as mentioned in Sec. II.B. It is not clear if the disagreement should raise a query on the consistency of Young's approach in straight geometry. The path difference causes no problems in Huygens-Fresnel diffraction theory; since the diffracted field is the sum of all the contribution from the unobstructed area.

A continuous phase change of the diffracted wave $U_{na}$ across the geometrical edges of the narrow slit can be seen by drawing phasors, $U_0$, $U_{na}$, $U_{nL}^B$, and $U_{nR}^B$, in the complex plane, in spite of that $U_0$ disappears and the phase of the boundary diffraction wave from the edge on the right $U_{nR}^B$ (or that from the edge on the left $U_{nL}^B$) changes by $\pi$ at the beam-shadow boundary. However, the $\pi/4$ advance of the phase remains unexplained; the phase advance can be 0 to $\pi/2$ in this argument.

## V. CONCLUDING REMARKS

The application of ultrasound to the demonstration of Fresnel diffraction is quite old. The zones and Poisson's bright spot have been shown by a Schilling apparatus.[16] The more recent piezo-ceramic transducer has made ultrasound measurements less difficult to perform. Fundamentals of waves, such as interference and diffraction, have been introduced mostly in optics course. However, the present work suggests that those subjects could be demonstrated and interpreted better in acoustics than in optics when the advantage of phase measurements is effectively used. Babinet's principle is confirmed directly by drawing phasors on the complex plane. The phase measurements showed some aspects of diffraction which otherwise are difficult to observe directly with the intensity measurements. A cylindrical wave originating at the edge of the straight screen as predicted by Sommerfeld diffraction theory has been clearly demonstrated. Curious results are observed in the phase measurements. The phase of the wave passing through the centre of the narrow slit showed that the Cornu spiral is tilted π/4 to the real axis as predicted by Huygens-Fresnel diffraction theory (see Eq. (4) and the paragraphs which follow).[3,5,8] The results obtained behind a strip on the central axis, indicates a disagreement between the phase and the path viewed from Young's approach, discussed at the next to the last paragraph of Section IV, is puzzling and needs more investigation.

The method presented may also be used in demonstrating mathematics introducing the Euler's identity: $e^{i\theta} = \cos\theta + i\sin\theta$, the complex plane, and the vector. In application, the phase measurements may provide additional information in application of nondestructive evaluation of materials in industry.

The quality of the experimental results is high enough to allow comparison with theory. In practice, a simple remote control of the receiver would reduce the time considerably in the phase pattern measurement since human movement can cause some disturbance of the air and it takes several tens of seconds to a few minutes to settle. Two screens were used to form a slit in the present measurement; however, a single screen with a rectilinear slit would make alignment simple. Sets of the apparatus for different frequencies are recommended when more than two groups of students perform the measurements to avoid the interference of each other.

**ACKNOWLEDGEMENTS**

The author is grateful to Prof. T.A. King for reading the manuscript and valuable comments. The author would like to thank Prof. T. Fujiwara, Dr. S. Takagi and Dr. H. Fukuda for helpful discussions on the edge diffraction wave.

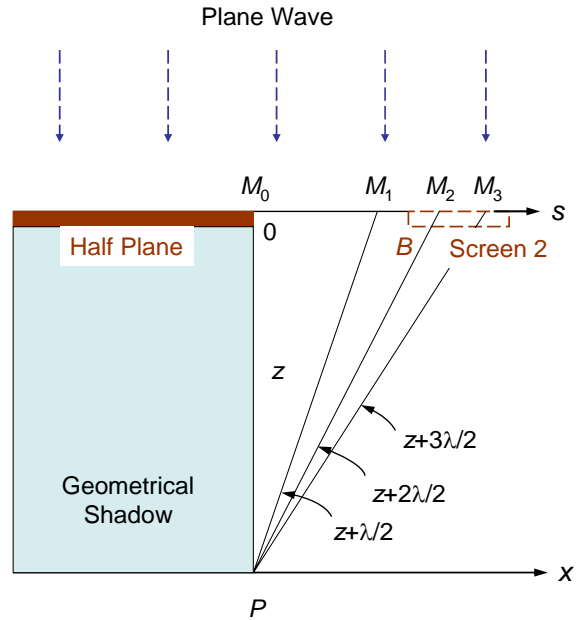

FIG. 1. Schematic view of the arrangement for a half plane with a straight edge shown in the *x-z* plane. The plane wave is incident normal to the half plane. The edges of the linear zones are shown ($M_i$). An additional screen (broken box) makes a slit.

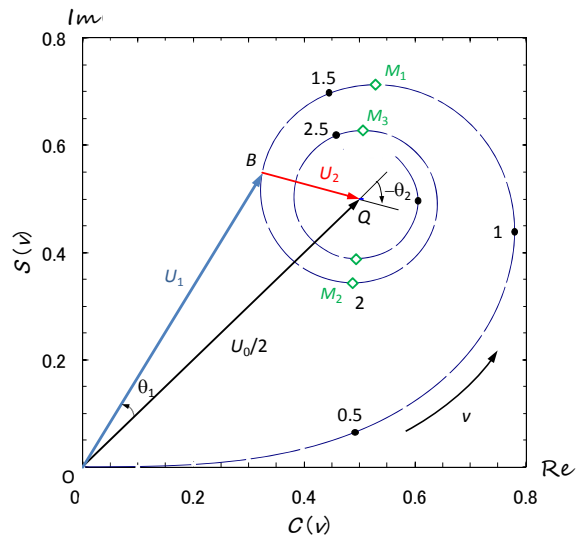

FIG. 2. The Cornu spiral for the first quadrant. The phasors $U_0/2$, $U_1$ and $U_2$ rotate clockwise with an angular velocity ω. See Fig. 7 for the direction of θ. The edges of the linear zones $M_i$ are shown by ◊. *v* values are also indicated on the spiral.

FIG. 3. Relation of phasors on the Cornu spiral for the observation point $P$ at $x=0$ (see Fig. 4). The straight line with arrow $I'J$ represents the phasor for a slit given by Eq. (4). Phasors for behind a half screen ($H'Q$; $H'$ moves on the Cornu spiral as $P$ moves on the $x$ axis) and a strip ($EQ$ and $QF'$) for $P$ off the axis ($x\neq 0$) are also shown with broken arrows.

FIG. 4. Typical configurations of two screens and a strip for the study of Babinet's principle. Thick horizontal lines show positions of screens and a strip placed on the $s$ axis in unit of $\lambda$ (Figs. 1 and 6). Positions for $v$ and the edges of zones $M_i$ are calculated for a plane incident wave for $z=10\ \lambda$. Letters $Q'$, $B'$, $O$, $B$ and, $Q$ corresponds to each position on the Cornu spiral shown in Fig. 3.

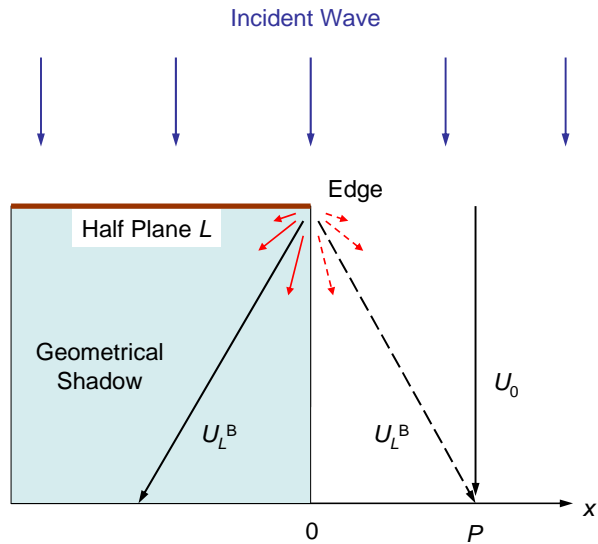

FIG. 5. The edge diffraction approach by Young. A cylindrical wave (short arrows) is produce at the edge of a half plane. The field observed in the geometrical shadow is this boundary diffraction wave $U_L^B$ and, in the direct beam, the interference between $U_L^B$ and the incident wave $U_0$. Broken arrows show a $\pi$ phase shift across the boundary of the geometrical shadow.

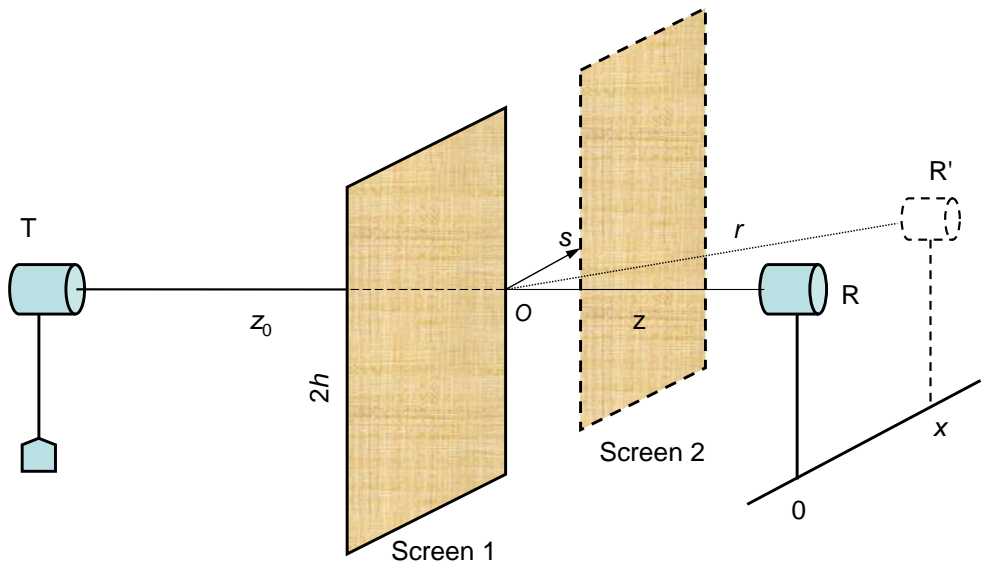

FIG.6. The schematics view of experimental setup for the study of the Cornu spiral using screens ($z_0, 2h \gg z > s, x$).

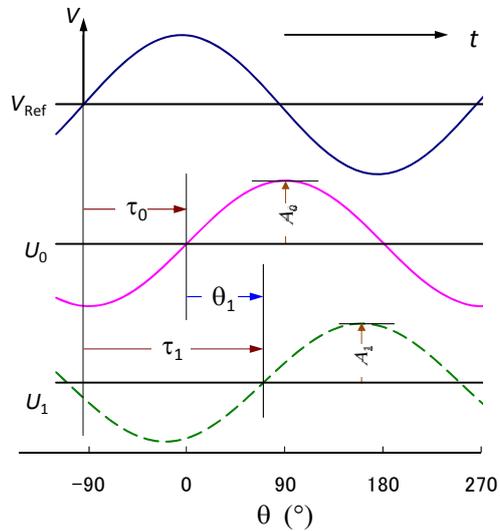

FIG. 7. The relation of waves observed by an oscilloscope between $U_0$ and $U_1$, and the reference voltage $V_{Ref}$ driving the transmitter. The phase $\theta$ is measured with respect to that of the incident wave $U_0$.

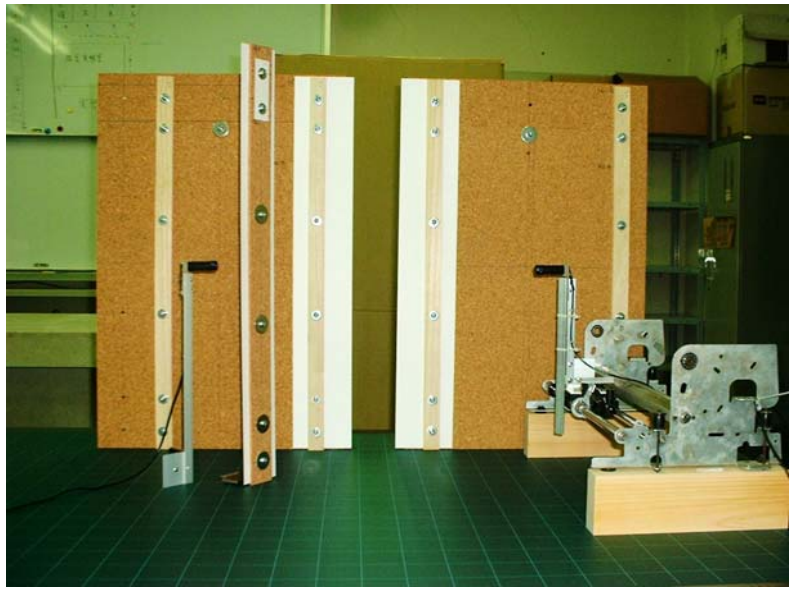

FIG. 8. Picture of the apparatus. From the left, the transmitter, the strip, and the receiver. Two screens forming a slit are shown at the back. The screens and strip are not aligned. The square-ruled mat provides a guide for alignment.

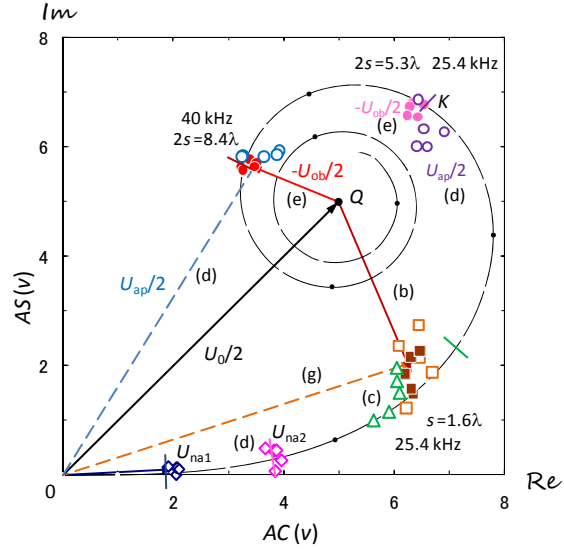

FIG. 9. The amplitude and phase relation for various sets of screens and strip observed at $x = 0$ (see Figs. 3 and 4). The amplitude $A_0$ of the incident wave is normalized to be $10\sqrt{2}$. The symbols are; ○: slits, ●: strips, △ and ■: screens, □: a slit, and ◊: narrow slits. The letters in brackets show the configuration as in Fig. 4 (positions $B'$ and $B$ are different). The bars on the Cornu spiral show the position of calculated $v$. $U_{na1}$ ($2s=\sim 1\lambda$) and $U_{na2}$ ($2s=2\lambda$) are the results obtained for narrow slits (see also Fig. 12).

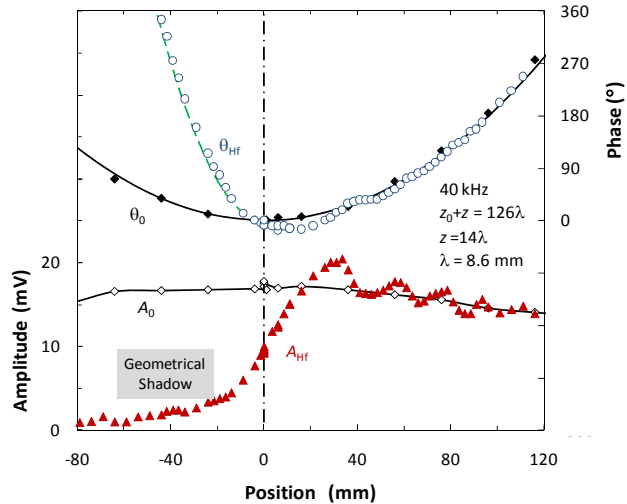

FIG. 10. The amplitude $A$ (bottom) and phase $\theta$ (top) observed for a half plane placed at $z = 120$ mm as a function of position $x$. The solid and broken parabolic curves show the phases calculated for a cylindrical wave originating at the transmitter and the edge of the screen, respectively.

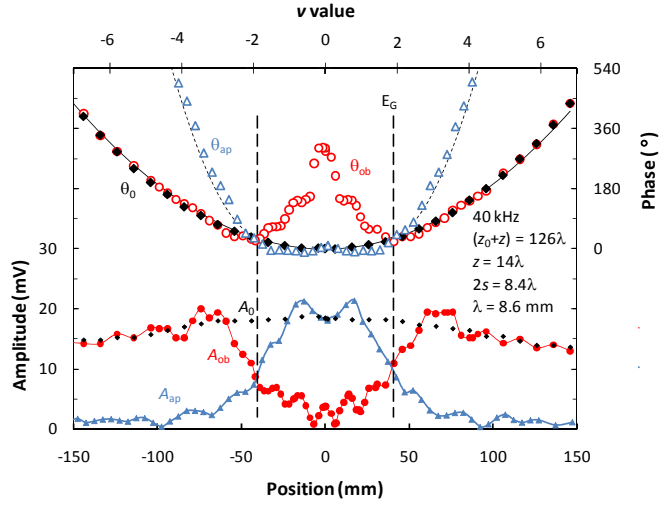

FIG. 11. The diffraction patterns for the amplitude $A$ (bottom) and the phase $\theta$ (top) as a function of the position $x$ observed for a slit and a strip of 72 mm width placed at $z = 120$ mm). The vertical broken lines show the edges of the geometrical shadow. The solid parabolic curve shows the phase of the cylindrical (spherical) wave originating at the transmitter. The broken curves show the cylindrical waves originating at the edges of the slit (see text).

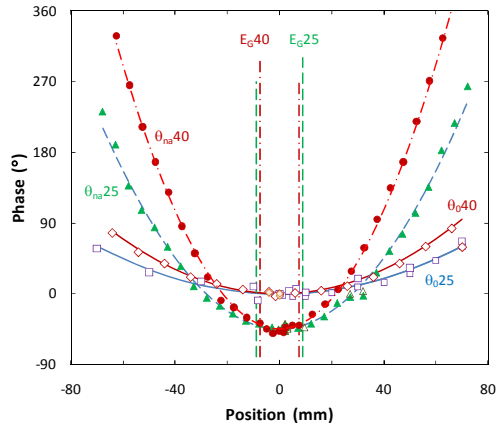

FIG. 12. The phase $\theta_{na}$ observed for narrow slits of about $1\lambda$ width as a function of $x$. The broken (25.4 kHz, $z = 135$ mm) and chain (40 kHz, $z = 120$ mm) curves show the phases of cylindrical waves originating at the centre of the slit with an added phase $\Delta\theta$ of -45° (see text). The vertical broken and chain lines ($E_G$) show the edges of the geometrical shadows. (See also $U_{na1}$ and $U_{na2}$ in Fig. 9.)